# Single spin resonance in a van der Waals embedded paramagnetic defect

*Nathan Chejanovsky * [1,2], Amlan Mukherjee [1], Youngwook Kim [2,3], Andrej Denisenko [1], Amit Finkler [1,4], Takashi Taniguchi [5], Kenji Watanabe [5], Durga Bhaktavatsala Rao Dasari * [1,2], Jurgen H. Smet [2] and Jörg Wrachtrup [1,2]*

[1] 3. Physikalisches Institut, Universität Stuttgart, Pfaffenwaldring 57, 70569 Stuttgart, Germany

[2] Max Planck Institute for Solid State Research, Heisenbergstr. 1, 70569 Stuttgart, Germany

[3] Department of Emerging Materials Science, DGIST, 333 Techno-Jungang-daero, Hyeonpung-Myun, Dalseong-Gun, Daegu, 42988 Korea

[4] Department of Chemical and Biological Physics, Weizmann Institute of Science, Rehovot 7610001, Israel

[5] National Institute for Materials Science, 1-1 Namiki, Tsukuba, 305-0044, Japan.

**Abstract**

Spins constitute a group of quantum objects forming a key resource in modern quantum technology. Two dimensional (2D) van der Waals materials are of fundamental interest for studying nanoscale magnetic phenomena. However, isolating singular paramagnetic spins in 2D systems is challenging. We report here on a quantum emitting source embedded within hexagonal boron nitride (h-BN) exhibiting optical magnetic resonance (ODMR). We extract an isotropic g-factor close to 2 and derive an upper bound for a zero field splitting (ZFS) ($\leq$ 4 MHz). Photoluminescence (PL) behavior under temperature cycling using different excitations is presented, assigning probable zero phonon lines (ZPLs) / phonon side band (PSBs) to emission peaks, compatible with h-BN's phonon density of states, indicating their intrinsic nature. Narrow and in-homogenous broadened ODMR lines differ significantly from monoatomic vacancy defect lines known in literature. We derive a hyperfine coupling of around 10 MHz. Its angular dependence indicates an unpaired electron in an out of plane π-orbital, probably originating from an additional substitutional carbon impurity or other low mass atom. We determine the spin relaxation time $T_1$ to be around 17 μs.

**Main Text**

Van der Waals two dimensional (2D) materials allow the isolation of one atomic monolayer to few layers of a crystal as thin as ~ 0.5 nm hosting a wealth of optically active defects. While a significant number of defects in various materials have been investigated, identifying their structural and chemical composition from optical spectra is challenging. Typically, other spectroscopic methods like, e.g. electron spin resonance (ESR) provide valuable additional insight. However, standard magnetic resonance is by far not sensitive enough to measure spin signals in 2D materials. Combining it with optical detection dramatically enhances sensitivity but requires the defect emission and absorption to depend on its spin state. In only a very few cases this has been demonstrated so far. In $WSe_2$, a narrow band-gap 2D material (1.35 eV [1]), quantum emitters (QE) originating from bound exciton quantum dots (QDs) whose emission wavelength is sensitive to large magnetic fields ( > ~ 1 Tesla) have been reported, [2,3] albeit lack spin dependent optical relaxation channels. Point defects with electronic states inside the band-gap can be highly localized with wavefunctions confined to the atomic scale, [4] exhibiting the strong exchange interaction necessary for spin dependent relaxation channels. On the other hand, QDs can have relatively large spread wavefunctions, due to confinement on a nano-scale range, encompassing thousands of atoms. [5] A prototypical host for optically active point defects is hexagonal boron nitride (h-BN), a graphene analog 2D Van der Waals material with a band-gap of 5.95 eV. [6] Various QE spanning a large emission wavelength range have been attributed to h-BN in the 2D [7,8,9,10] and 1D form [11] with their chemical structure not conclusively identified but rather computationally conjectured using DFT. [7,12,13] Recently, a h-BN QE with magnetic field dependent emission has been found. [14] Due to the large surface area/volume ratio in thin 2D crystals both sample preparation methods and source material have a huge impact on defect density. [11] For example, chemically functionalized 2D graphene, lacking a band-gap, can host room temperature QE placing them under the category of QDs [15] and *not under intra band-gap* point defects.

In the following we demonstrate that a paramagnetic emitter in h-BN [14] exhibits ODMR upon confocal intra-band excitation. We study the emitter's electron spin properties including hyperfine structure and spin relaxation time, PL behavior under cryogenic conditions and absorption polarization. From the data we debate the possible spin configurations and structure.

In Figure 1 we first analyze the paramagnetic emitter in terms of spatial location, PL, anti-bunching and magneto-optic response at room temperature followed by PL and absorption polarization at cryogenic conditions. We use single crystal h-BN as our source material from Ref. 16, which has become a hallmark for high quality h-BN in research. QE in h-BN suspended from a substrate emit with a reduced count rate, [17] therefore for increased collection efficiency our h-BN is placed on SiO$_2$ substrates, allowing us to collect photons at a maximum rate of 250 KHz using 633 nm excitation. (See SI). Probing a h-BN multilayer with dimensions of (~ 1 x 2 μm) using 3 linearly polarized excitation lasers (594/633/730 nm) reveals a bright emitter which is similarly spatially localized for all excitations roughly at the center of the flake (Figure 1.a). Reflection measurements and wide bright/dark field images (see SI) reveal the flake structure, hinting that the spatial location of our emitter is not in a flat region, but rather a bended/broken edge, similar to previous reports. [18]. The room temperature PL spectra using 594/633/730 nm excitations (Figure 1.b) reveal a number of emission peaks which we enumerate as P$_i$(T). Some peaks are enumerated together using the P$_{i+j}$ syntax as they overlap or, as will be shown in cryogenic conditions, are composed of two main peaks. According to Ref. 14 the room temperature ZPL of emitters that respond to magnetic fields is located at 727 nm which correspond to our peaks P$_{4+5 \text{ (594nm/633nm)}}^{(298K)}$ = 727 ± 18 / 729 ± 29 nm in Figure 1.b. A close observation in ref. [14] also reveals a small peak at 812 nm, hardly visible using our 594 nm excitation but easily identified using longer wavelengths excitations, P$_{7 \text{ (633nm/730nm)}}^{(298K)}$ = 812 ± 2 / 813 ± 4 nm. In brief, we surmise, that Peaks in the range of i >3 have spectral components of our paramagnetic emitter, whereas Peaks 0-2 most probably arise from a nearby defect(s). These components are analyzed in detail below. For more details see Figure 1 caption.

Looking into the confocal image of our emitter (Figure 1.c) compared to a different non-paramagnetic h-BN emitter of singular nature (marked with 'g$^2$(0) < 0.5'), reveals that the paramagnetic emitter has a larger spatial spread for all 3 excitations and has an additional 'bump' on the right side. This bump disappears only for 730 nm excitation. This spatial information suggests that even under 730 nm excitation we are not probing a single emitter. Next, we test the emitter's magneto-optic response to a magnetic field generated from a permanent magnet. A

Gaussian fitted count distribution (Figure 1.d), shows a non-over-lapping center peak, indicating a similar magneto-optic response, previously reported. [14] This magneto-optic response seen here as an emission reduction, was also observed as an increase in emission under different conditions (See below). Hanbury Brown and Twiss autocorrelation measurements ($g^2(t)$) can reveal the number of emitters involved in the emission process and can yield insights on the electronic structure, such as the existence of a metastable state. In Figure 1.e the red dashed lines of values '1' / '0.5' represent the threshold for bunching/single emission of the system for $|t| > \sim 10$ ns / $|t| < \sim 10$ ns, respectively . Our $g^2(t)$ function (for $|t| > \sim 10$ ns) is above 1, indicating the presence of a meta-stable state (bunching), which is of significance to observe ODMR. Applying / Removing the magnetic field (Figure 1.e. On/Off) modulates the bunching/anti-bunching behavior as the frequency of decay rates from the excited to ground and metastable states are changed, most prominently seen on the $\tau_2$ component, consistent with reduction of emission count (Figure 1.d). For $|t| < \sim 10$ ns the $g^2(t)$ function doesn't go below 0.5 and thus *more than one QE* dominates the emission process. Specifically, we calculate $\sim 3$ emitters. This is in agreement with our emitter spatial spread in Figure 1.c. Similarly, previous reports [14] have shown that $g^2(0)$ was not conclusively below 0.5. Thus one can speculate that this type of emitter tends to form in proximity to each other. [19]

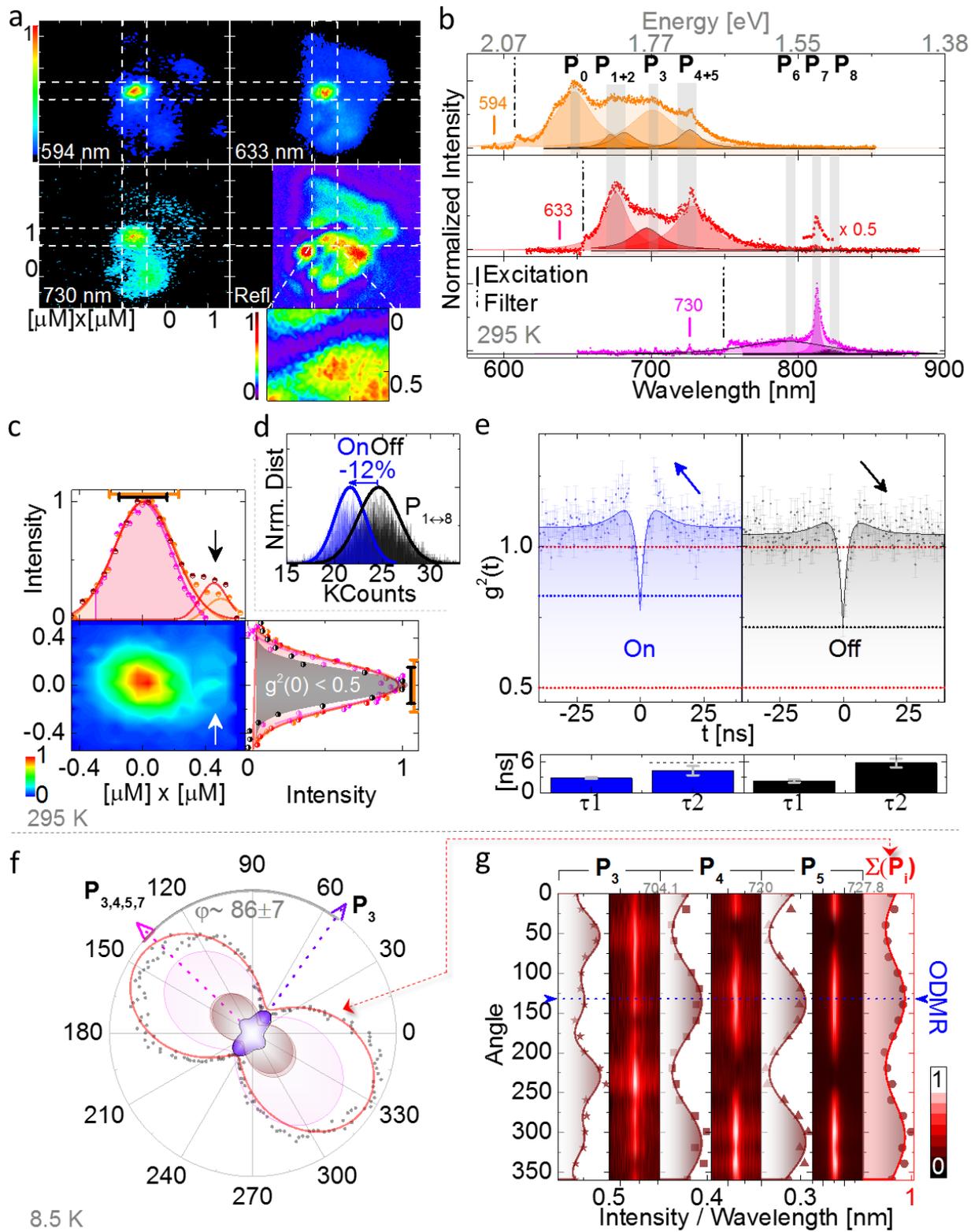

Figure 1: (a) 4 panels display the emitter under 594/633/730 nm and a reflection image using low power 532 nm excitation. The emitter is localized in the same spatial coordinates under all excitations. The reflection measurement reveals the emitter is in a spatial area containing perimeters. (b) Room temperature PL spectra using 3 linearly polarized excitations of 594nm / 633 nm /

*730 nm (top to bottom).Peak numbers are enumerated as $P_i (0 \leq i \leq 8)$ for each Lorentzian fit. Note that a Voigt fit yielded the best fit for 730 nm excitation. Excitation wavelength and optical filters used are marked in the graphs. Peak 7 is enlarged for clarity for 633 nm excitation. (c) High resolution confocal scan of the emitter for 594nm / 633nm / 730 nm excitation (orange, red, purple curves, respectively). The emitter is less localized in comparison to the confocal spread of another non-paramagnetic emitter of singular nature, (black bar comparison in x and y profiles). (d) Gaussian count distribution with/without a magnetic field (black/blue, respectively). The peaks contributing to the emission are marked. (e) Auto-correlation measurements with/without a magnetic field (black/blue, respectively). A modulation is more prominently seen in the $t_2$ component (f) Absorption polarization plot of the emitter at 8.5K excited with 633 nm. The grey dots are the counts collected from the APDs when rotating the excitation with a fitted function (red line). (g) PL spectra for all angles plotted per peaks 3/4/5 and sum of peak 3/4/5. The peak/sum designation is indicated on the top scale. Left side is the summed profile of the color bar intensity plot on the right for each peak wavelength range. Each tick in the color bath plot is 2 nm and the end wave length is indicated on the top scale. The summed profile was fitted with a diploe function for peaks 4/5 and the total and a quadrupole function for peak 3 as explained in the text. The summed profiles for peaks 3/4/5 are super imposed in (f) in magenta, dark red and purple, showing the misalignment of 86 degrees of some components of peak 3 from all the others. Blue arrows indicate the excitation angle used in the ODMR measurements below. Each peak's polarization is superimposed on the total emission count in f (rescaled for clarity).*

Rotating the linear excitation polarization can reveal information regarding emitter absorption polarization and orientation. In Figure 1.f/g we study this dependence by monitoring emission count rate and PL spectral features at cryogenic conditions of 8.5K. The emission intensity variation as a function of excitation angle (grey points) of the 633 nm laser is fitted using a $\sin^2(\theta - \theta_0)$ (Figure 1.g, red curve). As photon auto-correlation revealed in Figure 1.e, emission stems from a few emitters. To see how the different emission peaks of those emitters behave, we record the PL for a subset of excitation angles from maximum to minimum emission counts (See SI). [14] The intensity image per peak of these are plotted for all excitation angles Figure 1.g (right), with the intensity sum profile per angle (left). We first discuss the behavior of Peaks 4/**5** as we identify them later on as likely candidate for the zero phonon line of the emitter. The fit to angular behavior indicates emission is comprised of linearly polarized emitters, in agreement with h-BN QE reports. [7,8,20,14] Peak 4/**5** show similar periodicity ($\sim \sin^2(\theta - \theta_0)$). Remarkably, the few emitters responsible for PL are parallel. Peak 3 exhibits two periodicities fitted best with a quadrupole function ($\sim \cos^2(\sigma - \sigma_0)\sin^2(\theta - \theta_0)$ ). Thus peak 3 has one component aligned with Peaks 4-7 (for Peak 7, see SI) and another deviating with $\varphi = 86 \pm 7°$ from the other peaks. A possible interpretation is that peak 3 stems from a higher excited state of the emitter but with different orbital symmetry.

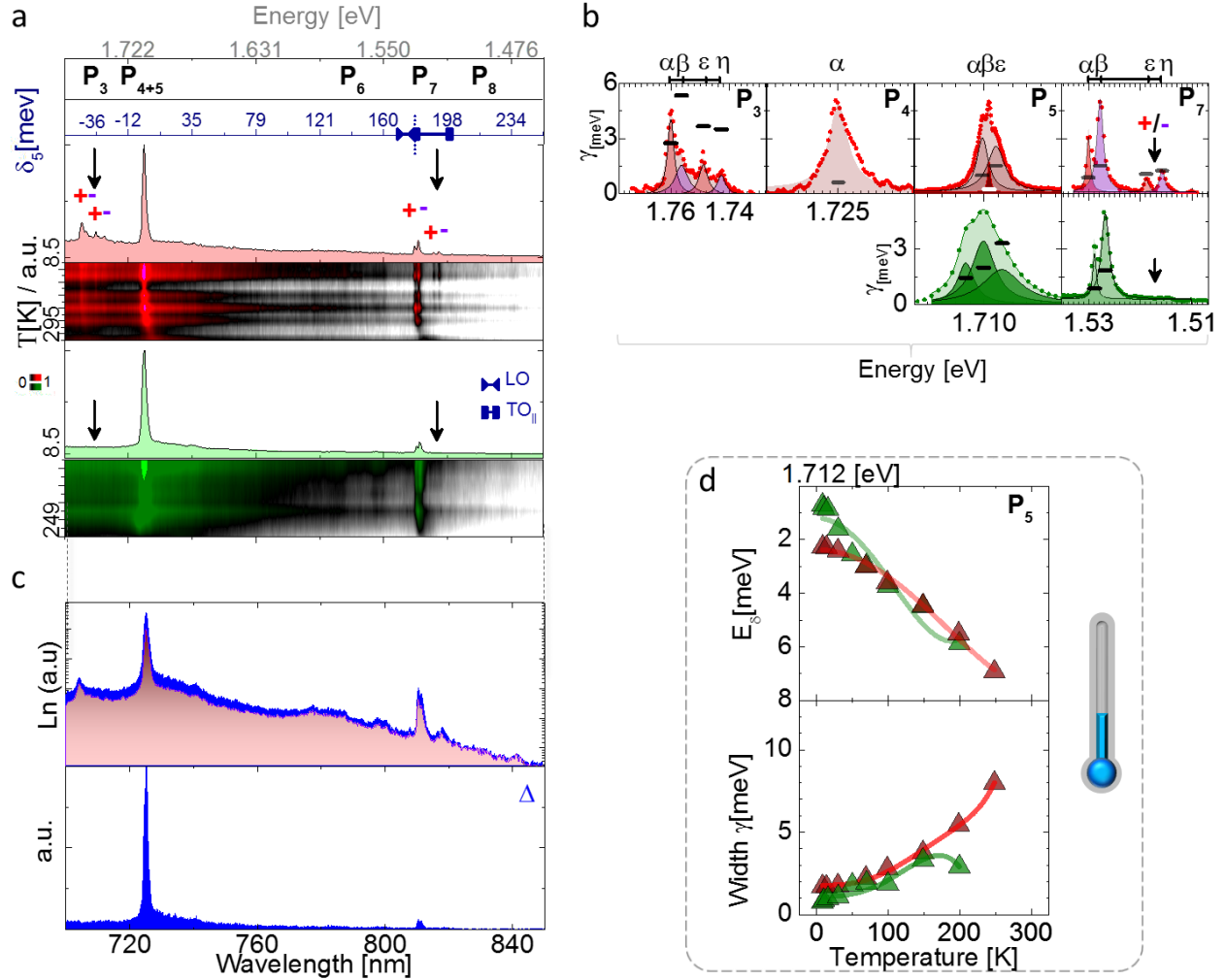

*Figure 2: (a) Temperature dependent intensity PL in a.u. units of the emitter using 633 nm/532 nm excitation at 8.5 K, red/green curves, respectively. The color bars underneath each represent the natural logarithmic scale of the PL intensity as a function of temperature. Note that the temperature scale is not linear but consists of the temperature points depicted in (e). The temperature outer boundaries are noted (8.5 K and 295 K, 8.5 and 249 K, 633 nm/532 nm excitation, respectively). Peak numbers are displayed in the upper frame. The dark blue scale represents the energy detuning of the spectra from peak 5 ($\delta_5$). The optical phonon energy range is displayed on the scale. Black arrow display features which are absent between both excitations (b) Normalized PL intensity for each peak at 8.5 K in (a), decomposed with Lorentzians, upper graphs represent excitation with 633 nm and bottom with 532 nm. As guide to the eye lines connecting components are displayed on top as they have similar energy spacing as described in the text. FWHW of components in (b) in ($\gamma$) meV are superimposed. (c) Top charts - PL spectra of the emitter with / without a magnetic field of 130 G (black/red curve, respectively). The difference in the PL spectra is the blue colored area, seen at the bottom. (d) Peak position shift (upper charts) and linewidth change (bottom charts) as a function of temperature for 633 nm excitation (red). See SI for fit analysis.*

While cryogenic PL measurements of various h-BN QE sources have been reported,[21,20,22,23] none address the cryogenic PL of a paramagnetic emitter. It has been proposed on two different h-BN QEs that if the energy difference between the excitation laser and the ZPL is above h-BN's maximum optical phonon energy, a Huang-Rhys (HR) two level electronic model is inadequate to explain the emission mechanisms. [24] For lower energy ZPLs it has been proposed that excitation is mediated by cross relaxation and not through direct laser excitation. [23] Different energy excitation can also shed light on possible charge states. [25,26] More importantly however we note, that we only see ODMR (see below) upon red (633 nm) excitation. Thus we display in Figure 2 the emitter's main PL features, line widths, and luminescent phonon behavior with 633/532 nm excitation separately (with color coding in red/green, respectively). PL a.u. intensity is shown in Figure 2.a with color bars of the natural logarithmic intensity scale for decreasing temperatures (bottom to top). In Figure 2b each peak is decomposed to components (denoted α,β,ε,η) using Lorentzians, revealing two peaks unresolved ($P_{4+5}$) at room temperature: $P_{4(633\ nm)}^{(8.5K)}$ = 719 and $P_{5(633\ nm)}^{(8.5K)}$ = 725 nm. Immediately noticeable are variations in the PL features for red/green excitations (black arrow markings). Peak 3/4 and components of peak 7 are absent using 532 nm excitation throughout the whole temperature range in Figure 2.a. possibly indicating different relaxation pathways. Nevertheless, peak **5** retains the largest intensity for both. To gain insight into the magneto-optic behavior, in order to reveal which peaks are affected by the magnetic field, we measure the PL and count rate distribution (See SI) of the emitter without/with a field of 130 G, red/blue curves, respectively. The comparison and Δ function of these reveal the dominate role of Peak **5** and lesser role of Peaks 3/4/7. In Figure 2.b the FWHM in meV (γ) (black lines) for each component are displayed. As most components have a width of < ~3 meV, smaller than h-BN bulk phonon DOS ( ~5 meV), [23,27] we cannot discard multiple ZPLs possibly stemming from multiple emitters. The brightest (for both excitations) and narrowest component for 633 nm excitation is Peak **5**,**β** ~ 0.25 meV (Figure 2.b) with a dominate role in the magnetic response. Therefore, we tentatively assign Peak **5** as a ZPL. To explain the role of the other peaks in Figure 2.b we plot the absolute value of detuning in energy of all other components to Peak **5** in Figure

2.a $\delta_5$ as a detuning scale in dark blue. Typically, h-BN QEs display a well separated PSB shifted from the ZPL, corresponding to ~169-200 meV. [8,11,23,24,17,28] Additionally, ungapped low energy acoustic phonons can cause multi-PSBs obscuring the ZPL.[23] Taking these into account fits well with our observations. The adjacent peaks of **5**/α/ε (Figure 2.b) are acoustic phonons detuned by ~ < 5 meV, obscuring the ZPL (**5β**). Peak 7 falls in the range of the stokes LO and TO optical phonons (superimposed blue double arrows markings in Figure 2.a), for both excitations and for the full temperature range (color bars, Figure 2.a). A more detailed phonon map highlights the dominant LO mode and ZA, LA, TO phonons (See SI). All these are clear fingerprints and evidence of the *emitter's coupling to the h-BN lattice*. The doubling of the spectral features for Peak 3 and 7 (with 633 nm excitation) seems to indicate an intermittent Stark shift effect which would shift the ZPL, obscured by acoustic phonons, which would translate to a shift in the optical phonons possibly due to a charge state switching, seen as the doubling of the LO mode in Figure 2.a/b.

In Figure 2.d, we analyze the temperature behavior for peak **5** in terms of position/FWHM change, ($E_\delta$/Width, upper/bottom, respectively). For simplicity, the analysis is valid for the first peak component in Figure 2.b. We adopt the ZPL power shift behavior known for defects in diamond [29,30] of $\sim a_\delta T^2 + b_\delta T^4$ and for the FWHM a $\sim a_\gamma T^3 + b_\gamma T^7$ behavior [23,21] with a possible $c_\gamma T^5$ contribution. [31] The emission energies are shifted to larger energies (blue shift) in the temperature range investigated. We tend to discard band-gap variations as the sole cause of the ZPL shift. In short, we surmise that the peaks 3/5/7 are correlated. See SI for a detailed explanation and for the temperature dependence of the other enumerated peaks.

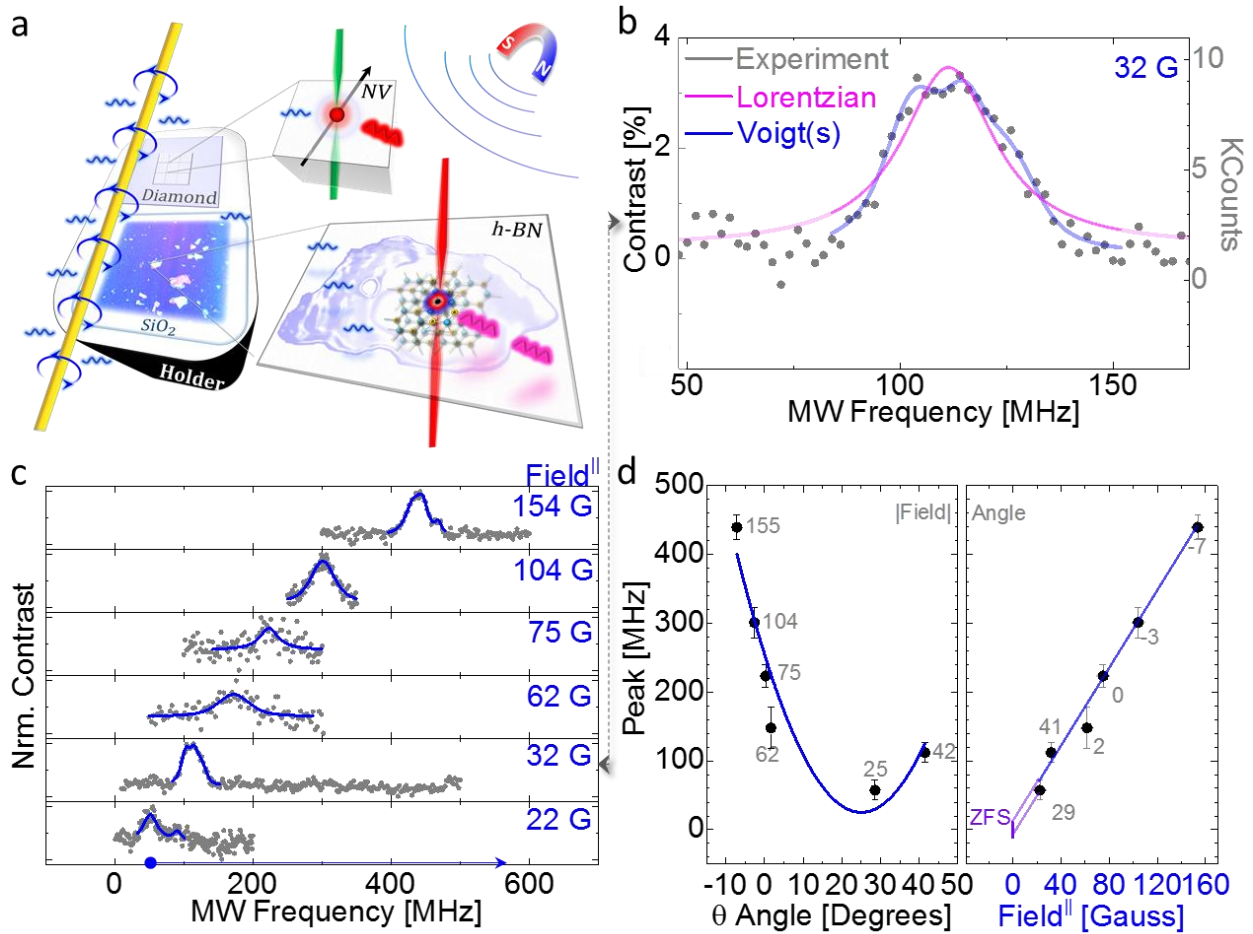

Figure 3: (a) A schematic of the configuration used to apply ODMR. A 20 μm conducting copper wire is spanned in proximity to a NV and our emitter in order to supply microwaves. The emitter is simultaneously excited with a continuous wave (CW) 633 nm laser.(b) Enlarged ODMR signal of the 32 G magnetic field measurement. Contrast/Net Counts are seen on the left/right axis, respectively. (c) Normalized ODMR contrast of MW sweeps for the h-BN emitter at different fixed magnetic field values. Parallel magnetic field value are indicated in blue. The ODMR MW frequency increases as the magnetic field is increased. Each peak is fitted using one or more Voigt functions. As the ODMR spectra can contain elements arising from inhomogeneous/homogenous broadening (Gaussian/ Lorentzian transitions signature, respectively), we fit the entire ODMR signal with a Voigt envelope which is a convolution of both. Some ODMR peaks contained asymmetric features and required more than one Voigt envelope to achieve a good total envelope fit. (d) Central envelope frequencies from (c) plotted as a function of magnetic field angle (left) and as a function of magnetic field strength (right). Grey values correspond to the magnetic field strength/angle.

We now turn to the observation of ODMR of the emitter, displayed in Figure 3. We note that ODMR was only seen at cryogenic condition. Microwaves (MW) for ODMR are applied by a thin wire (20 micro meter thick) in close proximity to the emitter. A magnetic field generated by a permanent magnet exterior to the cryostat is changed in strength/orientation by displacing the magnet. To yield quantitative information on the local magnetic field strength/orientation a diamond with single NV⁻ defects is placed in close proximity to the BN defects. All experiments are carried out at 8.5 K. The measuring scheme is depicted in Figure 3.a. See SI for the equations used to determine the magnetic field/angle. Upon recording the fluorescence intensity of the emitters, we sweep the MWs in the region where we expect the spin resonance of a free electron (g-factor = 2) to occur and indeed register an increased luminescence with a maximum change of the photoluminescence of the defects by ~ 5%. Figure 3.b displays an example of such a measurement with a 3% contrast. When changing the magnetic field strength, we observe the expected Zeeman shift of the resonance line (see Figure 3.c). However, we do not see any systematic changes in the line width or shape when changing the magnetic field strength (without changing its angle). We discuss this further below. The behavior of the resonance packet central frequency and width is described by a spin Hamiltonian comprising the interaction of a single electron with N nuclear spins:

$$H = g_e \beta B_0 S + SDS + \sum_{i=1}^{N} g_n \beta B_0 I_i + SAI_i + I_i Q I_i.$$

Here $g_{e,n}$ is the electron/nuclear g-factor, $B_0$ the external magnetic field, S and I are the electron /nuclear spin operators, D the fine structure tensor, A the hyperfine tensor and Q marks the nuclear quadrupole interaction. Extrapolating the slope/intercept in Figure 3.d (right) with the frequency axis we extract $g_e$ (slope) of 2.06, close to the free electron value. Our $B_0$ calibration procedure also allows us to determine the angle of the magnetic field and variation of the angle yields the $g_e$ factor anisotropy (see Figure 3.d – left). The resonance line and hence the g-value follows the following relation: $g = g_\perp + \nabla g \cos^2\theta$, where $\Delta g = g_\parallel - g_\perp$. Within error bars we find $\Delta g=0$, i.e. an isotropic g-factor. Only very few data on g-factor anisotropy are available in literature. Moore et al. [32] find a $\Delta g = 9.5 \times 10^{-4}$, i.e. only a small isotropy which would be below our detection accuracy. $g_e$ is of similar magnitude to that of paramagnetic defects in h-BN measured via EPR. [33]

g-Factors between 2.0024 to 2.0032 are reported in literature with larger g-factor tendency for more carbon rich materials owing to the larger spin-orbit coupling of C compared to B. Reducing $B_0$ to zero results in a measurement of the zero field splitting parameter D. As apparent from Figure 3.d (right) our measurements yield a value of $|D| \leq 4$ MHz with an uncertainty given by the fit to a linear Zeeman shift. In addition, we do not detect any ODMR line without applying a magnetic field nor did we find any indication of a second transition frequency. As a result, our measurements do not allow us to conclusively conclude on the spin multiplicity of the electron spin sublevels we probe. We discuss this further below.

Next, we probe the line-width shape of the spin resonance transition. Already Figure 3.c shows that there is no marked increase in the FWHM upon increasing the magnetic field. Thus we conclude the line-width is not governed by inhomogeneous broadening, e.g. a dispersion of g factors or orientations among the few defect centers we are probing which would cause a line broadening upon magnetic field increase. Rather we conclude that the line width and shape is dominated by unresolved hyperfine coupling (hfc). Indeed, most measured ODMR lines are in-homogenously broadened and some cases even seem to be asymmetric (see Figure 3.c). In the following we will discuss our line shape analysis. We note that an unresolved hyperfine and quadrupolar coupling, given by the last two terms in the spin Hamiltonian in general does not lead to a well justified determination of the chemical and structural composition of the defect. This is why we restrict our analysis to the comparison to a few known spin defects in h-BN. We go through those before analyzing our results. For the simple case of a paramagnetic mono-vacancy ($N_V$ or $B_V$) center, EPR of $N_V$ is more abundant in the literature. Two hyperfine structures have been observed for the $N_V$ in h-BN: An unpaired electron coupled to three nearest neighbor B nuclei (three boron center –TBC) and an unpaired electron coupled to one adjacent B nucleus (one boron

center – OBC). [33,34,35] For the OBC, typically oxidative damage forces the electron to interact only with a single B nucleus. For both centers, the g-factor and hyperfine parameters have been measured. The g-factors measured are all in the range of g = 2 and hence agree within error bars to our measurement, whereas hfc parameters show a significant variance. Specifically, the OBC shows an axially symmetric hfc tensor with main elements $A_{\parallel/\perp}$ = 247.8 / 326.2 MHz. The TBC has shown two sets of hfc values $A_{\parallel/\perp}$ = 112 /127.8 MHz and $A_{\parallel/\perp}$ = 18.4 / 22 MHz. We use these parameters to compare simulations using the above Hamiltonian to our results (See below, Figure 4.a). The entire pristine h-BN lattice exhibits nuclei with magnetic momentum. The most abundant isotopes of nitrogen and boron being $^{11}B$ [80%], I=3/2 and $^{10}B$ [20%], I=3. Both nuclei show a quadrupolar moment which we neglect for the present analysis. Their gyromagnetic ratios differ by a factor of 3, but this does not show up in the spectra as we assume only allowed electron spin resonance transitions with $\Delta m_I$=0. More importantly the ODMR resonance line shape and width will depend on the number N of nuclei the electron is coupling to. For N nuclei of identical hfc parameter A, the number of equally spaced lines scales as 2NI + 1. With I=3 for $^{10}B$ this results in 19 equally spaced lines composing the spectrum in Figure 3.b/c. Alternatively, an electron spin coupled to only a single $^{11}B$ would result in only 4 hyperfine components.

In Figure 4.a we compare simulations with those hyperfine parameters with our experimental results. A depiction of each simulated center is shown on the left. The stick spectra mark the frequency position of individual spectral components. For a more quantitative comparison we convolute the component linewidth in form of a Lorentzian function ($C_N$' – right side). We assume that hfc of the electron spin only occurs with boron nuclear spins at their natural isotope composition and that the magnetic field is aligned along the $A_\parallel$ direction. First of all, the simulated spectra differ drastically with the clearest deviation being the OBC with four ($^{11}B$) or 7 ($^{10}B$ hyperfine lines separated by about 300 MHz. The TBC simulation also clearly deviates from the measured spectrum with an overall linewidth of more than a factor of ~ 5. The carbon related nitrogen vacancy (marked as $C_N$) is the TBC which shows the smallest hfc. However, it also is too large to fit our data. Only when we further reduce the hfc to ~ 6 MHz (marked as $C_N^{'}$) the simulation fits our measured ODMR line. We note that in our analysis we discard other ways to simulate the ODMR spectrum, for example an OBC with different hfc than those reported in the EPR literature or coupling to nitrogen nuclei.

Assuming that a single electron spin coupled to three boron nuclei is the most rational fit to our experimental data we now explore the angular dependence of the ODMR linewidth. Our simulations allow for a qualitative insight of the hfc tensor. The measured hyperfine couple should depend as $A(\theta) = A_\parallel + \Delta A cos^2(\theta)\ with\ \nabla A = A_\perp - A_\parallel$. Hence a measurement of the linewidth as function of θ yields insight into ∆A and hence $A_\perp$. Figure 4.b. shows a comparison of experimental ODMR data at two different magnetic field angles θ with the corresponding simulation. The experiment yields a decrease in line width of around 30%, from 41.3 MHz for θ = 0° to 31.2 MHz for θ = 40°. This is best simulated by using $A_{\parallel/\perp}$=10 / 6.8 MHz. Assuming this to be the optimum hfc parameters we derive at an optimum fit of the ODMR line (see Fig. 4a), assuming a homogeneous line width of around 7 MHz equivalent to a $T_2$ time of around ~ 0.2 μs. This is two orders of magnitude shorter than that calculated for the hypothetical $N_BV_N$ point defect, [36] probably arising from residual neighboring electron spins or further unresolved hfc in our analysis.

Further insight into the spin dynamics of the defect is gained upon analyzing the relaxation times T1 and T2$^*$. We assume a hfc coupling to one $^{11}$B and determine these using the equation states in the SI. We extract from the power dependent contrast fittings and the Lorentzian FWHW component of the Voigt fittings $T_1 = 17 \pm 4$ µs and $T_2^* = 57 \pm 10$ ns (See SI). The FWHW and contrast power dependence are in agreement with reports for NV ODMR. [37] ODMR at room temperature was not seen, possibly due to rapid spin-lattice relaxation time ($T_1$) which improves upon cooling down to cryogenic temperatures. [33] The similar value of $T_1$ to that reported in EPR studies is another footmark of a possible carbon defect. [38]

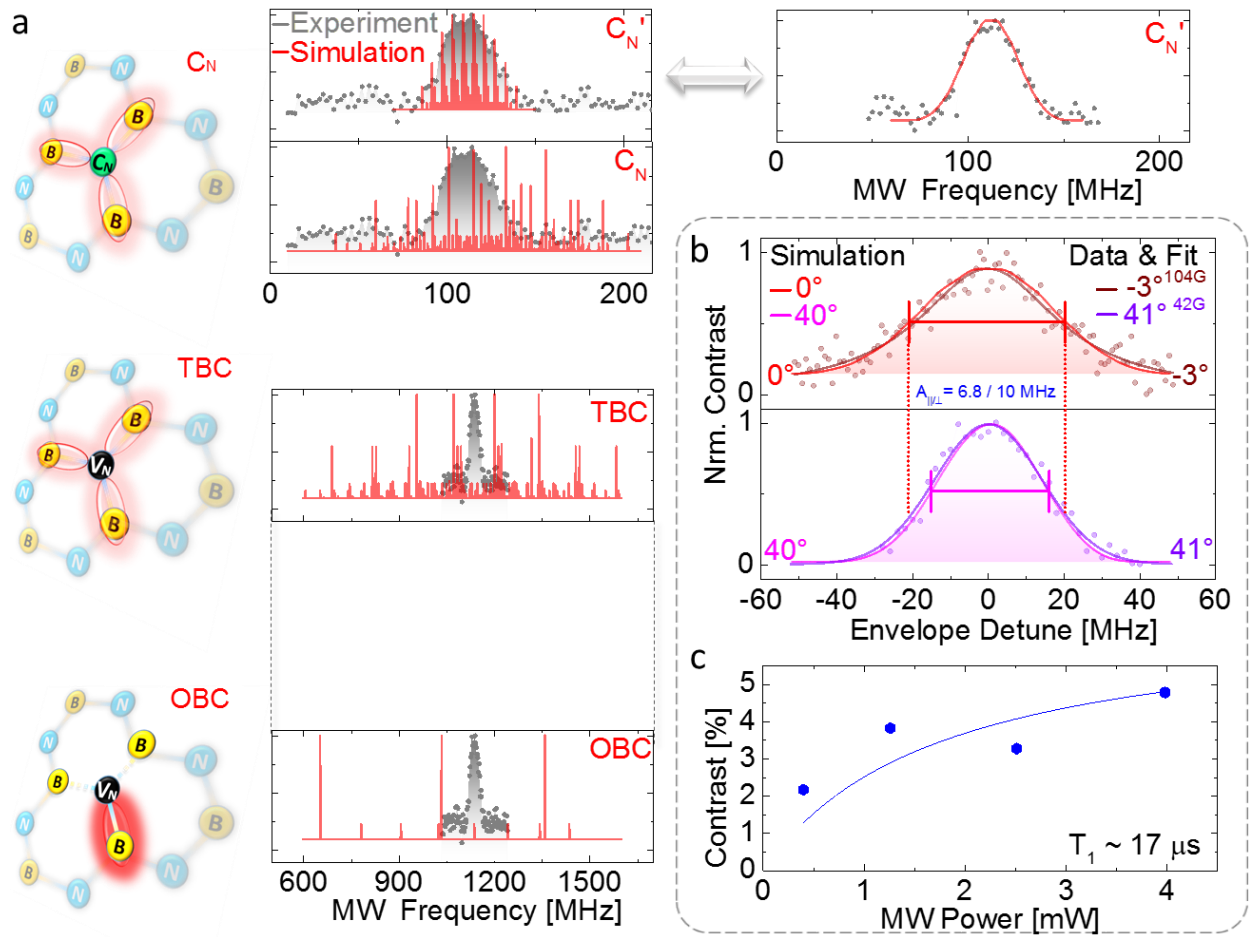

Figure 4: (a) Schematics of three point defects in h-BN, a carbon substituting a nitrogen atom and a nitrogen vacancy, hyperfine coupled to three near boron atoms (first two) and one boron atom (last). The simulated ODMR spectra for the configuration of each of these is displayed in red on the right superimposed on the experimental measurement (in grey). The relatively larger FWHM of the TBC point defect rules it out as possible candidate for the observed para-magnetic emitter. A carbon substitute atom fits under the experimental envelope of the ODMR signal. (b) Simulating the tilting of the magnetic out-of-plane angle results in variations of the ODMR envelope, with a good fit to the data envelope, confirming our interpretation of unresolved hyperfine components in our ODMR signal.(c) Superposition of all ODMR power dependent measurements with a decomposition assuming a hyperfine coupling of 13.5 MHz to 1 boron atom. Power dependent measurements for a fixed magnetic field, for which we extract $T_1$.

We now reflect on the data to gain insight on the atomic electronic structure. Literature EPR has given insight into the structure and electron density of the $N_V$ defect in h-BN. The only defects shown to exhibit similar narrow EPR lines are those found upon binding carbon atoms to the $N_V$, which also stabilizes the defect. Essentially, the carbon atom drags away spin density from the boron atoms and hence reduces hfc. The magnitude and ratio of $A_\parallel$ to $A_\perp$ further gives insight into the dominant wavefunction contribution responsible for hfc. The unpaired electrons occupy an atomic orbital $\psi = C_S \phi_S + C_P \phi_P$ where $\phi_S$ and $\phi_P$ are the 2s and 2p orbitals. f and d are further calculated as $f \propto |C_S|^2 \eta |\phi_S(0)|^2$ and $d \propto |C_P|^2 \eta <\phi_P(0)|1/r^3|\phi_P(0)>$, where $\eta$ is the total spin density at the atom. The values $|\phi_S(0)|^2$ and $|\phi_P(0)|^2$ are distinct for B and N and are calculated from ab initio calculations of the atoms. For the in-plane $sp^2$ σ-orbitals of h-BN, the expected values for the coefficients are $|C_S|^2 = 1/3$ and $|C_P|^2 = 2/3$, i.e. the σ-orbitals result in both contact and dipolar interaction. In contrast the out of plane p orbitals yielding $|C_S|^2 = 0$ and $|C_P|^2 = 1$, that is only yield dipolar interaction. Upon using standard values from $|\phi_S(0)|^2$ and $<\phi_P(0)|1/r^3|\phi_P(0)>$ one can calculate hyperfine parameters for the σ and π orbitals. Taking into account that the spin density η is spread over multiple orbitals in h-BN one yields an hfc of ~ 30 MHz for π-orbitals and ~ 300 MHz for σ-orbitals. The observed hfc thus points towards a major contribution from out of plane π-orbitals. This is also consistent with our simulation of the anisotropy of the hfc. If the average $\langle 1 - 3cos^2(\Phi)\rangle$ is taken over an out of plane π-orbital the resulting ratio of $A_\parallel$ to $A_\perp$ is 2, close to what is observed in the experiment. The ODMR *increased* photon emission is also known for defects in silicon carbide [39], hinting to an initial preferential spin polarization of the dark state in the h-BN emitter, modified to the bright state by the MW, as opposed to the NV⁻ which is preferentially spin polarized in the bright state. Calculations on the possible spin configurations in ref. [14] have suggested that a spin system (S) of S > ½ should be capable to exhibit ODMR. Such a system should have a ZFS which we have not observed in our measurements with the possibility that the ZFS was smaller than the hfc. For higher order systems (S ≥ 1) one or more satellite ODMR peaks should appear, which we have not observed. More detailed measurements should be conducted in the future for verification. A literature survey reveals that our narrowest ODMR line width of 31 MHz (Figure 3.c) is in the same order ( < ~100 MHz) to those seen in other EPR studies.[34,40] Interestingly for carbon and hydrogen/oxygen defects in h-BN a similar EPR line width has been observed *also with unresolved* hyperfine interaction. [33] For carbon, hyperfine splitting can be eliminated due to rapid electron exchange of nitrogen vacancies and nearby carbon(s). This can be caused by disorder, resulting from a high lattice defect concentration, [32] or reduction in the electronic delocalization on neighboring boron atoms. [33] This agreement between our ODMR observations and the EPR suggests that the defect structure consists not only of vacancy defects but also *additive impurity atoms* which could be introduced during exfoliation or thermal annealing, [11,32] similar to DFT predictions of point defects such as the $V_NC_B$ defect. [12,13] More experiments while carefully varying the magnetic field angle might render the hyperfine satellites resolvable as seen in EPR. [32] Assuming a low frequency ZFS we can also explain the decrease and increase of our emission rate when a magnetic field is applied. Intrinsic h-BN ferromagnetism due to edge N-termination [41] or $N_V$ defects in h-BN with long-range magnetic interaction [42] possibly generate an intrinsic magnetic field which the external applied field must first overcome to observe 'zero' field emission rate.

In summary we demonstrate for the first time ODMR on single or few defects in a two dimensional van der Waals material. We displayed the detailed non-resonant PL features of the emitter at 8.5K at various temperature cycling and polarization absorption properties. Our results give further insights on the electronic structure and phonon coupling of the emitter which can help to pin point in more precision its exact chemical nature. Our observation of a low energy peak at > 800 nm (peak 7) can assist in future research to identify or isolate more emitters with these properties as the emission in this wavelength is known to be scarce for QE in h-BN. [20]

**Corresponding Author(s)**

* Inquiries should be sent to the following e-mail address:

natan.chejanovsky@pi3.uni-stuttgart.de and d.dasari@pi3.uni-stuttgart.de

**Author Contributions**

All authors contributed to the paper. N.C., J.W. and A.F. conceived and designed the experiments. T.T. and K.W. grew the single crystal h-BN material. Y.K. and N.C. prepared the h-BN samples. A.D. fabricated the NV implanted diamond. N.C. and A.M. performed all measurements. N.C., D.D and J.W. analyzed the data. J.W. and J.H.S. supervised the project. N.C. wrote the manuscript with input from J.W., D.D., A.F. and J.H.S.

**Acknowledgments**

The authors acknowledge support from the Max Planck Society as well as the EU via the project DIADEMS and the DFG. J.H.S. acknowledges financial support from the EU graphene flagship. K.W. and T.T. acknowledge support from the Elemental Strategy Initiative conducted by the MEXT, Japan and JSPS KAKENHI Grant Numbers JP26248061, JP15K21722 and JP25106006.

**ABBREVIATIONS**

h-BN, hexagonal boron nitride; ZPL, Zero phonon line; QE, quantum emitters; PL, photo-luminescence; APD, avalanche photo diode; CW, continuous wave; FWHM, full width half maximum; TO/LO phonons, Transverse/Longitudinal optical phonons; ISC, inter system crossing; DOS, density of states; hfc, hyperfine coupling .